\documentclass[twocolumn,twoside,slac_two]{revtex4}
\usepackage{graphicx}
\usepackage{fancyhdr}
\usepackage{aas_macros,url}

\newcommand{\sA}{\sigma}
\newcommand{\sigv}{\langle \sA v \rangle}

\begin{document}

\title{The VERITAS Dark Matter Program}

%

\author{Alex Geringer-Sameth\footnote{Electronic address: \url{alex_geringer-sameth@brown.edu}} for the VERITAS Collaboration}
\affiliation{Department of Physics, Brown University, 182 Hope St., Providence, RI 02912}

\begin{abstract}
The VERITAS array of Cherenkov telescopes, designed for the detection of gamma-rays in the 100 GeV-10 TeV energy range, performs dark matter searches over a wide variety of targets. VERITAS continues to carry out focused observations of dwarf spheroidal galaxies in the Local Group, of the Milky Way galactic center, and of Fermi-LAT unidentified sources. This report presents our extensive observations of these targets, new statistical techniques, and current constraints on dark matter particle physics derived from these observations.
\end{abstract}

\maketitle

\thispagestyle{fancy}

\section{Introduction}

The characterization of dark matter beyond its gravitational interactions is currently a central task of modern particle physics. A generic and well-motivated dark matter candidate is a weakly interacting massive particle (WIMP). Such particles have masses in the GeV-TeV range and may interact with the Standard Model through the weak force. Searches for WIMPs are performed at particle accelerators \citep{2007PhR...438....1F}, where dark matter may be produced in high-energy collisions, and by low-background direct detection experiments which look for the scattering of dark matter particles off nuclei \citep{2004ARNPS..54..315G}. In astrophysics, the problem is being addressed through the field of indirect detection \citep{1996PhR...267..195J,2000RPPh...63..793B}. Here the goal is the detection of the end-products of dark matter annihilation which can take place throughout the Universe.

Dark matter annihilation into Standard Model particles generically gives rise to high-energy gamma-rays. There are several classes of experiments currently searching for hints of such gamma-ray emission. The Large Area Telescope (LAT) onboard the Fermi Gamma-ray Space Telescope constantly surveys the entire sky at energies of  about 100 MeV up to a few hundred GeV. At higher energies, ground-based atmospheric Cherenkov telescopes perform searches for emission from specific targets. In this contribution I review the dark matter searches currently being carried out by the Very Energetic Radiation Imaging Telescope Array System (VERITAS).

\section{VERITAS}
The VERITAS array \citep{2002APh....17..221W,2006APh....25..391H} consists of four 12-meter imaging atmospheric Cherenkov telescopes (ACTs) located at the base of Mount Hopkins at the Fred Lawrence Whipple Observatory in Arizona. Such telescopes use the Earth's atmosphere as a target for high-energy cosmic particles. An incoming gamma-ray may interact in the Earth's atmosphere, initiating a shower of secondary particles that travel at speeds greater than the local (in air) speed of light. This entails the emission of ultraviolet Cherenkov radiation. The four telescopes of the VERITAS array capture images of the shower using this Cherenkov light. The images are analyzed to reconstruct the direction of the original particle as well as its energy.

VERITAS is sensitive to showers initiated by gamma-rays with energies from around 100 GeV to more than 30 TeV. Because of the steep energy spectra of typical sources, fluxes in the VERITAS energy band are much smaller than those seen with Fermi. However, the difficulty of measuring such small fluxes is mitigated by the enormous effective areas of ACTs. VERITAS is sensitive to gamma-ray showers incident over an (energy-dependent) area of approximately $10^9\,\mathrm{cm^2}$ (compare with $\sim10^4 \,\mathrm{cm^2}$ for the LAT). At 1 TeV the energy resolution of VERITAS is approximately 15\%. The array has an angular resolution of about $0.14^\circ$ at 200 GeV and $0.1^\circ$ at 1 TeV.

\section{Dark matter targets}

The particle physics governing the annihilation of cold dark matter is the same everywhere. Therefore, any location in the Universe that hosts a suitably high dark matter density is a target for an indirect search. The flux of gamma-rays from dark matter annihilation takes a surprisingly simple form
\begin{equation}
\frac{dF}{dEd\Omega}(E,\theta) = \frac{\sigv}{8 \pi M_\chi^2} \frac{dN_\gamma(E)}{dE} \frac{dJ(\theta)}{d\Omega}.
\label{eqn:flux}
\end{equation}
Here, $dF/dEd\Omega$ is the flux of gamma-rays per energy per solid angle, $\sigv$ is the velocity-averaged cross section for dark matter self-annihilation and $M_\chi$ is the mass of the dark matter particle. The quantity $dN_\gamma / dE$ is the energy spectrum of gamma-rays arising from a single annihilation. The last factor quantifies the distribution of dark matter along the line of sight. The so-called $J$ value is defined as
\begin{equation}
\frac{dJ}{d\Omega} = \int dl \rho_\chi^2,
\label{eqn:dJdO}
\end{equation}
where $\rho_\chi$ is the dark matter density and the integral is taken along a particular line of sight.

Note that the dependence on the environment is completely captured in the $J$ value; the other quantities are universal properties of the dark matter particle. In particular, the energy spectrum of annihilation is governed by the annihilation channel (e.g annihilation into quarks, leptons, photons, etc). Only at cosmological distances is this spectrum expected to be attenuated by absorption or pair-production.

Unlike charged particles, gamma-rays travel undeflected from their point of emission. The excellent angular resolution of VERITAS therefore makes it a powerful instrument because it can target specific locations with a high dark matter density (large $J$ value). The Collaboration is currently focused on four classes of dark matter targets: nearby dwarf galaxies, the Milky Way galactic center, nearby galaxy clusters, and unidentified sources in the Fermi catalog.

\subsection{Dwarf galaxies}

Milky Way dwarf spheroidal galaxies are dark-matter dominated systems that orbit inside the gravitational potential of the Galaxy \citep{2012arXiv1205.0311W}. They are excellent targets for indirect detection because they are nearby, have large dark matter densities, and perhaps most importantly, contain no known gamma-ray sources. Therefore, any detection of gamma-rays from this class of objects becomes intriguing evidence for the annihilation of dark matter.

In the absence of a signal, constraints can be placed on the dark matter mass and annihilation cross section so long as one has an estimate of the $J$ value. Currently, the dark matter distribution in the dwarfs is modeled using the motions of their member stars (e.g. \cite{2007PhRvD..75h3526S,2008ApJ...678..614S}).

\begin{figure}
\includegraphics{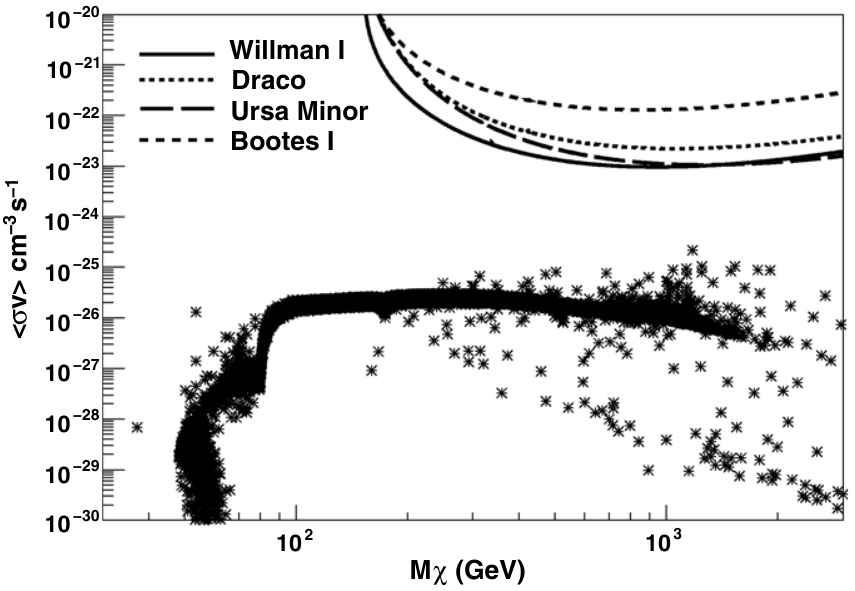}%
\caption{\label{fig:4dwarfs} Upper limits on the dark matter annihilation cross section derived from VERITAS observations of four Milky Way dwarf galaxies. Asterisks correspond to a model scan of MSSM parameter space subject to a relic abundance constraint. Figure taken from \citet{2010ApJ...720.1174A}.}
\end{figure}

VERITAS has published the results of observations of Bo{\"o}tes I (14 hr), Draco (18 hr), Ursa Minor (19 hr), and Wilman 1 (14 hr) \citep{2010ApJ...720.1174A} and deep observations of Segue 1 (48 hr) \citep{2012PhRvD..85f2001A}. Figure~\ref{fig:4dwarfs} shows the upper limits on the dark matter annihilation cross section as a function of dark matter mass derived from observations of the first four of these dwarf galaxies. The energy spectrum of gamma-rays from an annihilation, $dN_\gamma/dE$, was derived from a representative set of MSSM parameters. The asterisks in the figure represent a scan over MSSM parameter space with the constraint that each model predicts a dark matter candidate with a relic density within 3 standard deviations of that determined by three-year WMAP observations \citep{2007ApJS..170..377S}. 

\begin{figure}
\includegraphics{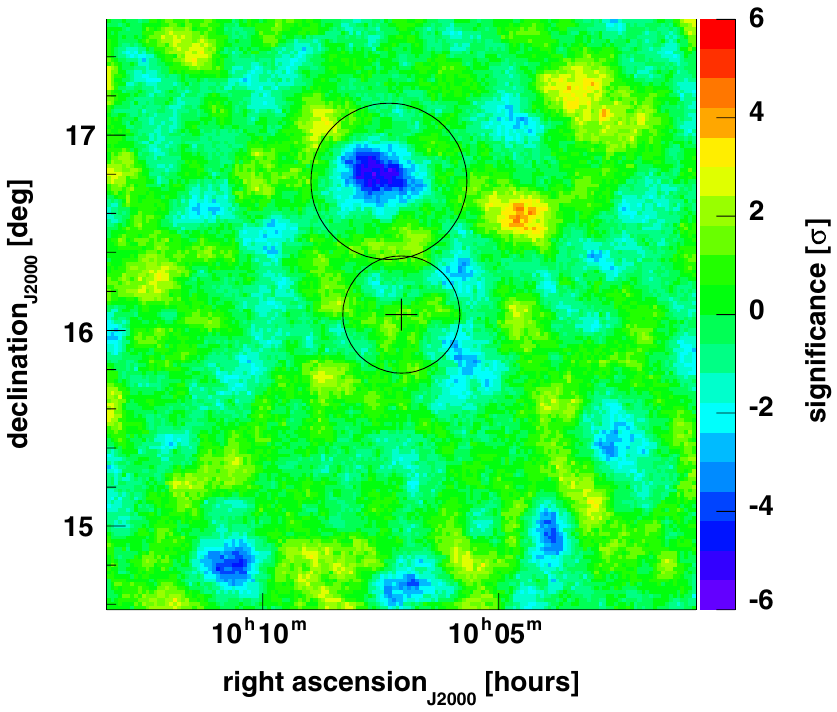}%
\caption{\label{fig:segmap}Significance map of the region surrounding dwarf galaxy Segue 1. The cross is at the location of the dwarf galaxy and the circles denote regions excluded when determining the background (one centered on Segue 1, the other on a bright star). Figure taken from \citet{2012PhRvD..85f2001A}.}
\end{figure}
No significant gamma-ray excess was detected from the 48 hour observation of Segue 1. The significance map of the region surrounding the dwarf galaxy is shown in Fig.~\ref{fig:segmap}. Here, the cross marks the location of Segue 1 and the circles denote regions excluded when determining the background. One is centered on the dwarf galaxy, the other on an unrelated bright star.
\begin{figure*}
\includegraphics{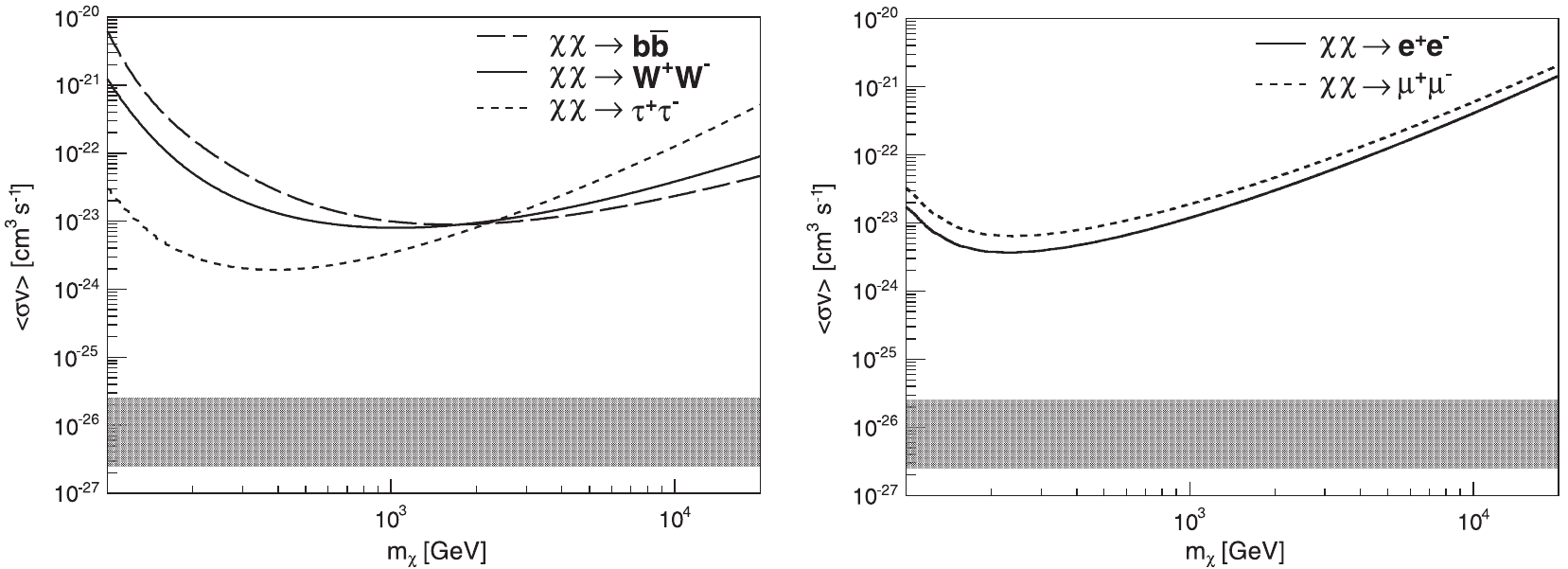}%
\caption{\label{fig:seg1lims} Annihilation cross section upper limits derived from observations of the dwarf galaxy Segue 1. Each curve represents a different annihilation channel. The cross section required to reproduce the observed relic abundance is shown by the gray bands. Figures taken from \citet{2012PhRvD..85f2001A}.}
\end{figure*}
As with the other dwarf galaxies, upper limits can be placed on the dark matter annihilation cross section. These limits are shown in Fig.~\ref{fig:seg1lims}, where each curve represents the limit from a different annihilation channel. The gray region indicates the cross section required to reproduce the dark matter abundance observed today. This represents a lower limit to the allowed cross section in generic WIMP models: WIMPs with a smaller cross section would be overabundant today. Therefore, the parameter space for general WIMP models is bounded at both ends. Further observations by VERITAS and many other experiments will continue to tighten the upper limits on the cross section.

\subsection{Clusters}

While further away, galaxy clusters are much larger than dwarf galaxies. In fact, these systems are the largest dark matter structures in the Universe. Because of their high dark matter densities and dense substructures they represent an attractive class of indirect detection targets. However, clusters contain large reservoirs of hot gas which will interact with cosmic rays. This results in the production of pions which then decay into gamma-rays. This gamma-ray emission, highly interesting in its own right, constitutes a background for dark matter searches.  In such searches there is often a tradeoff between high dark matter densities and astrophysical backgrounds.

\begin{figure}
\includegraphics{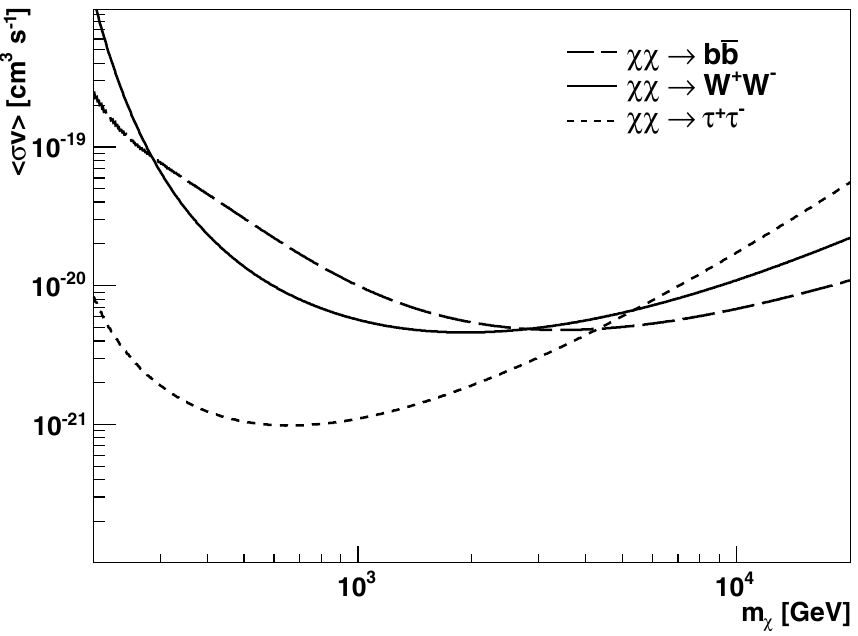}%
\caption{\label{fig:comalims}Upper limits on the dark matter annihilation cross section derived from observations of the Coma galaxy cluster. Different lines are upper limits assuming different annihilation channels. Figure taken from \citet{2012ApJ...757..123A}.}
\end{figure}

VERITAS has published comprehensive results from a 19 hour observation of the Coma cluster \citep{2012ApJ...757..123A}. No significant gamma-ray excess was seen. This allows interesting constraints to be placed on models of Coma's cosmic ray population and magnetic fields as well as on dark matter annihilation. Figure~\ref{fig:comalims} shows the upper limits on the dark matter annihilation cross section derived from the null-detection of Coma. In this analysis no gamma-ray background from cosmic rays was assumed, making the limits conservative.

\subsection{Galactic center}

The Milky Way galactic center is expected to be, by far, the brightest source of dark matter annihilation. This is due to its proximity (8 kpc) and its very large expected dark matter density. Because baryons dominate the mass distribution in this region there is a great deal of uncertainty on the dark matter distribution near the Galactic center. Nonetheless, the $J$ value for the galactic center is likely several orders of magnitude larger than those of the nearby dwarfs. The search at the Galactic center is complicated by a very large astrophysical background of gamma-rays arising from known and unknown point sources and from cosmic ray interactions with gas.

\begin{figure}
\includegraphics[width=8cm]{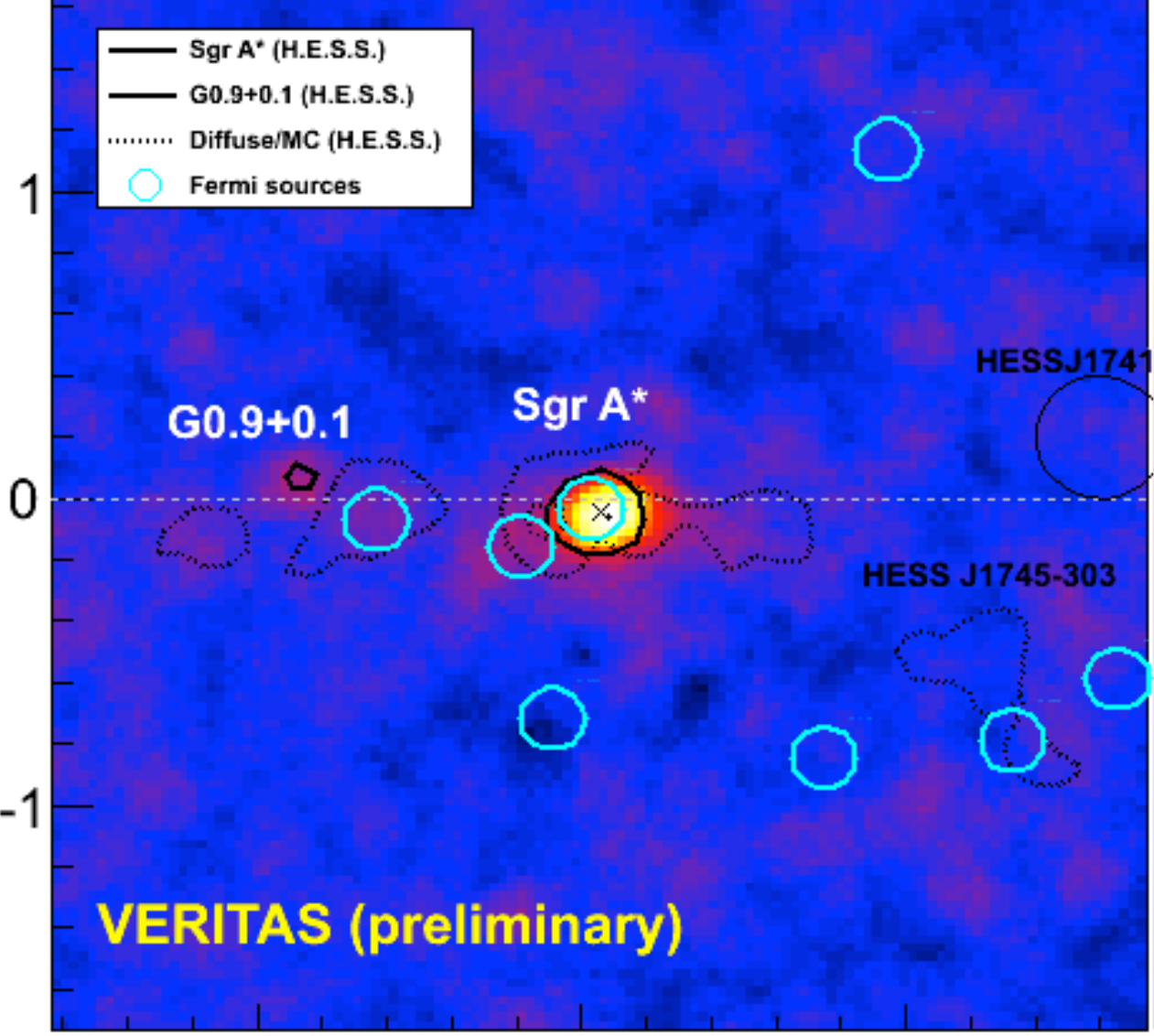}%
\caption{\label{fig:gcmap}Preliminary significance map of the Galactic center based on 46 hours of observation with VERITAS. Sources from the Fermi catalog are shown as light blue circles. Sources and diffuse emission detected by H.E.S.S. are shown with solid and dotted black contours.}
\end{figure}

Figure~\ref{fig:gcmap} shows a preliminary significance map derived from 46 hours of VERITAS observations of the Galactic center. The locations of sources in the Fermi catalog \cite{2012ApJS..199...31N} are shown with light blue circles. Sources detected by H.E.S.S. as well as diffuse emission detected by H.E.S.S. \citep{2006Natur.439..695A} are shown with black solid and dotted contours.
Emission from the galactic center is detected at 19$\sigma$. However, it cannot be unambiguously claimed that this signal to due to dark matter annihilation. The H.E.S.S. atmospheric Cherenkov telescope has reported the detection of diffuse emission along the Galactic ridge. Dark matter is not expected to cluster in the Galactic plane and so this diffuse emission is almost certainly the result of other astrophysical processes. The expected spherical morphology of the dark matter emission, however, can be used in an indirect search. The idea is to look toward the Galactic center but slightly away from the Galactic plane, avoiding much of the astrophysical background. This analysis is ongoing within the VERITAS collaboration.

\subsection{Fermi unidentified objects}
The Fermi LAT has detected many gamma-ray sources that have no known counterpart at other wavelengths. It has been suggested that such unidentified objects may be nearby dark matter halos. To explore this possibility two of these sources were selected for observation by VERITAS. The selection criteria are based on how likely the source is to be astrophysical as well as on the detection prospects for VERITAS. Variable sources as well as those close to the Galactic plane are ruled out as they are not likely to be dark matter halos. Sources with hard spectra and those detected at the highest energies by the LAT are preferred.
\begin{figure*}
\includegraphics[width=8cm]{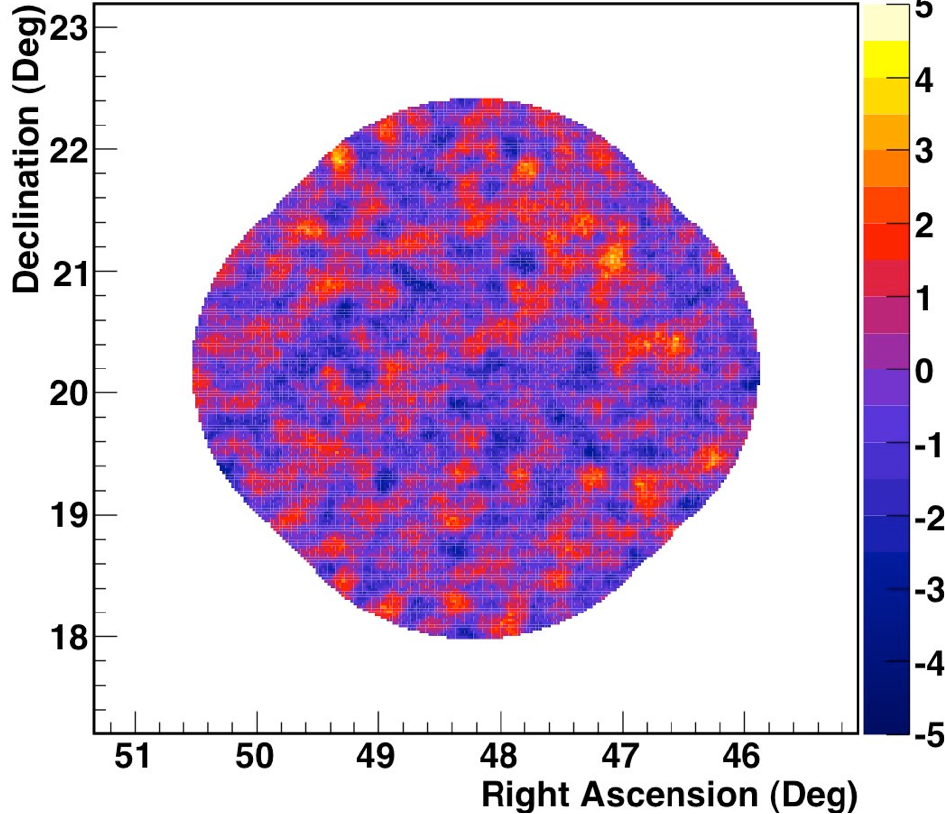}
\includegraphics[width=8cm]{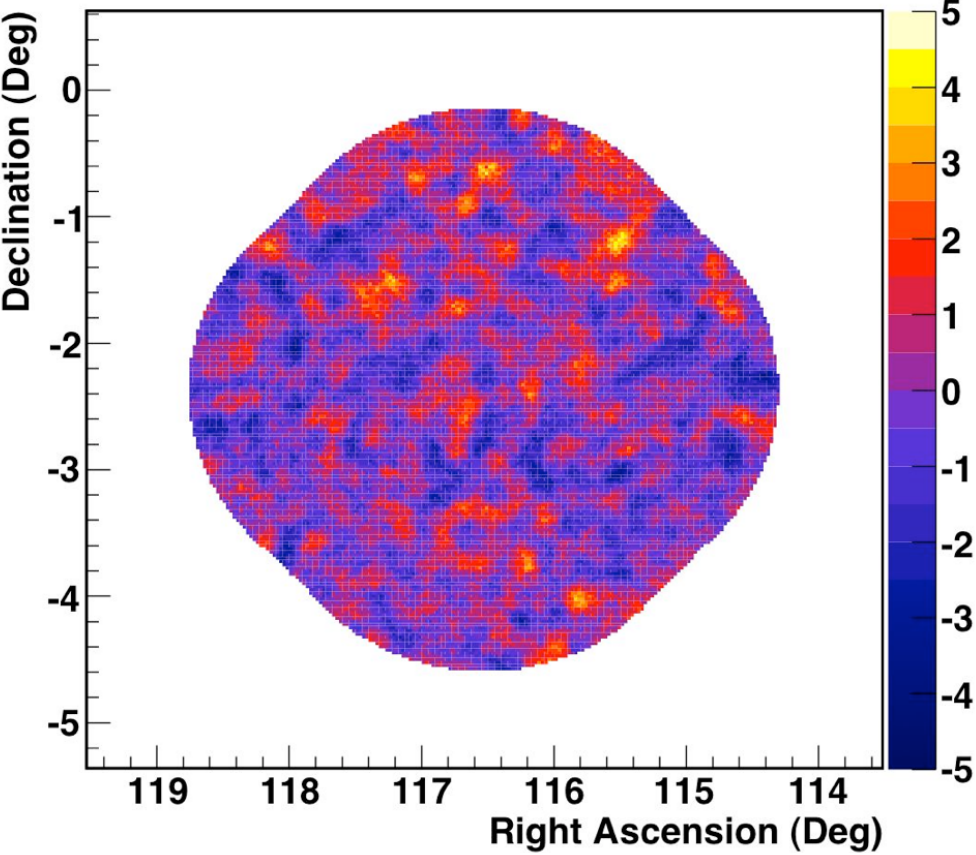}
\caption{\label{fig:ufos}Preliminary significance map of two Fermi unidentified sources observed with VERITAS. {\em Left:} 2FGL J0312.8+2013. {\em Right:} 2FGL J0746.0-0222.}
\end{figure*}
Figure~\ref{fig:ufos} shows significance maps for the two sources which have been observed. They are 2FGL J0312.8+2013 (10 hr) and 2FGL J0746.0-0222 (9 hr). No significant detection can be reported from either at this time.

\section{New analysis techniques}
Along with deeper observations of multiple dark matter targets, the VERITAS collaboration is exploring new analysis techniques designed to optimally extract the particle physics from the observations. So far, all dark matter limits have been constructed separately for each individual target. However, the physics of dark matter annihilation is the same across all targets. Therefore, it makes sense to combine the data taken on multiple observations into a single dark matter result. This is being performed using the framework developed in \citet{2011PhRvL.107x1303G,2012PhRvD..86b1302G}. In this analysis, detected events from different fields are weighted according to how likely they are to be due to dark matter as opposed to background processes (i.e. a Segue 1 event is given more weight than a Bo{\"o}tes I event since Segue 1 has a larger $J$ value). This will allow the complete collection of observations to be reduced into a single dark matter search or cross section upper limit.

\section{Conclusions}

VERITAS continues to perform dark matter searches across multiple classes of targets. Significant additional exposure has already been devoted to Draco (35 hr), Ursa Minor (36 hours), and Segue 1 (48 hours) with more observations planned. This will result in a data set of nearly 200 hours of observation across several dwarf galaxies. The full data set will be analyzed simultaneously, resulting in the most sensitive possible search given the available data. The VERITAS collaboration is proceeding with observations and analysis of the Galactic center and of Fermi unidentified sources and will publish its results in the near future.

\bigskip 
\begin{acknowledgments}
This research is supported by grants from the U.S. Department of Energy Office of Science, the U.S. National Science Foundation and the Smithsonian Institution, by NSERC in Canada, by Science Foundation Ireland (SFI 10/RFP/AST2748) and by STFC in the U.K. We acknowledge the excellent work of the technical support staff at the Fred Lawrence Whipple Observatory and at the collaborating institutions in the construction and operation of the instrument. The National Radio Astronomy Observatory is a facility of the National Science Foundation operated under cooperative agreement by Associated Universities, Inc.

I would like to thank Savvas Koushiappas, Ben Zitzer, Karen Byrum, Bob Wagner, Andy Smith, Jim Buckley, Jamie Holder, Martin Pohl, and Ken Ragan for their support in preparing this presentation. AGS is supported by a Galkin Fellowship at Brown University.
\end{acknowledgments}

\bigskip 
\bibliography{manuscript}

\end{document}